\documentclass[aps,prb,twocolumn]{revtex4}

\usepackage{graphicx}
\usepackage{dcolumn}
\usepackage{amsmath}
\usepackage{amssymb}
\usepackage{wasysym}
\usepackage{color}

\usepackage{mathtools}

\begin{document}

\title{Hidden valence transition in URu$_2$Si$_2$?
}

\author{N.~Harrison and M.~Jaime
}

\affiliation{Mail~Stop~E536,~Los~Alamos~National Labs.,Los~Alamos,~NM~ 87545
}
\date{\today}

\begin{abstract}
The term ``hidden order'' refers to an as yet unidentified form of broken-symmetry order parameter that is presumed to exist in the strongly correlated electron system URu$_2$Si$_2$ on the basis of the reported similarity of the heat capacity at its phase transition at $T_{\rm o}\approx$~17~K to that produced by Bardeen-Cooper-Schrieffer (BCS) mean field theory. Here we show that the phase boundary in URu$_2$Si$_2$ has the elliptical form expected for an entropy-driven phase transition, as has been shown to accompany a change in valence. We show one characteristic feature of such a transition is that the ratio of the critical magnetic field to the critical temperature is defined solely in terms of the effective quasiparticle {\it g}-factor, which we find to be in quantitative agreement with prior {\it g}-factor measurements. We further find the anomaly in the heat capacity at $T_{\rm o}$ to be significantly sharper than a BCS phase transition, and, once quasiparticle excitations across the hybridization gap are taken into consideration, loses its resemblance to a second order phase transition. Our findings imply that a change in valence dominates the thermodynamics of the phase boundary in URu$_2$Si$_2$, and eclipses any significant contribution to the thermodynamics from a hidden order parameter. \end{abstract}
\maketitle

\section{Introduction}

URu$_2$Si$_2$ remains of immense interest owing to the possibility of it exhibiting a form of broken-symmetry distinct from that observed in any other known material.\cite{mydosh2011} The term ``hidden order" has been coined\cite{kasuya1997,chandra2002} to describe reports of a BCS-like phase transition at $T_{\rm o}\approx$~17~K,\cite{palstra1985,maple1986} yet the absence of any signatures of symmetry-breaking reconcilable with the change in entropy at the transition.\cite{mydosh2011} Proposed forms for the as-yet-undetected symmetry-breaking fall roughly into two classes. In one of these, the $5f$-electrons are regarded as being itinerant, with the hidden order either leading to the opening of a gap on a pre-existing heavy Fermi surface\cite{chandra2002,okazaki2011,varma2006,elgazzar2009,fujimoto2011,pepin2011} or playing an integral part in the formation of the heavy Fermi liquid state itself.\cite{chandra2013,dubi2011} In the other, the $f$-electrons are regarded as being localized in an $5f^2$ configuration, with the hidden order involving interactions between local dipolar or multipolar degrees of freedom.\cite{santini1994,haule2009,ikeda1998,ikeda2012,kiss2005,harima2010,ohkawa1999,barzykin1995,kusunose2011,kung2015} Relatively little consideration has been given, however, to the possible consequences of the intermediate valence character of the $5f$ electrons in URu$_2$Si$_2$. Electron energy-loss spectroscopy\cite{jeffries2010} and resonant x-ray emission spectroscopy\cite{booth2016} indicate that a non-integer number of between 2.6 and 2.9 $5f$ electrons are confined to the atomic core, suggesting that a description of the $f$ electrons in terms of {\it either} itinerant {\it or} localized states is likely to be an oversimplification. 

\begin{figure}[!!!!!!!htbp]
\centering 
\includegraphics*[width=.45\textwidth]{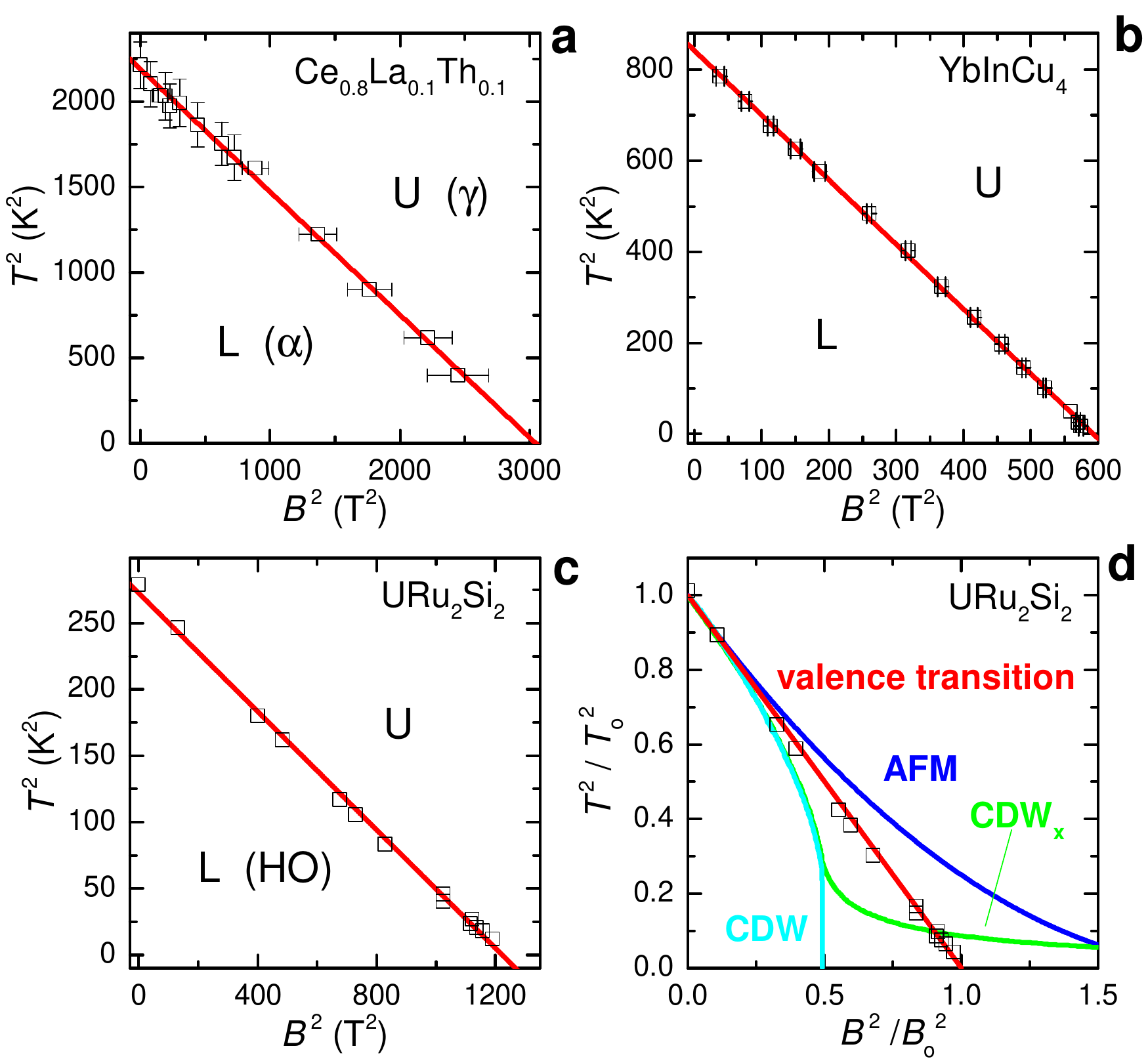}
\caption{{\bf a} - {\bf c}, Temperature-versus-magnetic field phase boundaries in Ce$_{0.8}$La$_{0.1}$Th$_{0.1}$,\cite{drymiotis2005} YbInCu$_4$\cite{immer1997} and URu$_2$Si$_2$\cite{jaime2002,kim2003} plotted in $T^2$-versus-$B^2$ coordinates. In URu$_2$Si$_2$, data points indicate the peak in the heat capacity at the transition\cite{jaime2002,kim2003} while $B$ refers to the magnetic field applied along the crystalline $\hat{c}$ axis. We neglect the additional phase transitions at $B\gtrsim$~36~T.\cite{kim2003i} {\bf d}, Comparison of the valence transition boundary with those of a charge-density wave (CDW),\cite{maki1964} an inhomogeneous CDW (CDW$_x$)\cite{mckenzie1997} and an antiferromagnet (AFM).\cite{moll2017} The curves have been renormalized to have the same slope $\partial(T^2/T^2_{\rm o})/\partial(B^2/B^2_{\rm o})$ in the limit $B\rightarrow$~0.}
\label{elliptical}
\end{figure}

Here we make the surprising finding that the temperature-versus-magnetic field ($T$ versus $B$) phase boundary of URu$_2$Si$_2$ closely follows the ideal elliptical form 
\begin{equation}\label{elliptical}
\bigg(\frac{T}{T_o}\bigg)^2+\bigg(\frac{B}{B_o}\bigg)^2=1
\end{equation}
characteristic of an entropy-driven transition,\cite{dzero2000,immer1997,drymiotis2005} as is known to occur for a discontinuous change in valence. We further find that the ratio of the critical temperature to the critical field, $T_{\rm o}/B_{\rm o}$, depends solely on the effective quasiparticle {\it g}-factor, and that this is adhered to rather precisely in URu$_2$Si$_2$. We also show that evidence for a valence transition is contained in the shape of the transition obtained from heat capacity measurements,\cite{jaime2002,kim2003} which we show to be significantly sharper than a BCS transition. Once quasiparticle excitations across the gap formed from the hybridization between conduction and $f$ electron states\cite{schmidt2010,aynajian2010,park2012,butch2015} are taken into consideration, the heat capacity no longer closely resembles a second order phase transition. Given these findings, a deeper investigation into the extent and origin of the irreversible behavior reported in certain experimental quantities at or below $T_{\rm o}$\cite{park1998,tonegawa2014,tabata2014,schemm2015} is warranted.

Valence transitions have been studied extensively in $f$ electron systems as a function of temperature, pressure and chemical substitution,\cite{immer1997,lawrence1981} with characteristic experimental features including an intermediate valence, a  discontinuity in the volume, a first order phase transition, and the typical absence of ordering at the transition. Owing to the high $T\gtrsim$~100~K temperature scale associated with archetypal $\gamma$-Ce to $\alpha$-Ce symmetry-preserving valence transition in pure cerium metal,\cite{lawrence1981} strong magnetic fields have made a relatively late contribution to our understanding of valence transitions.\cite{immer1997,drymiotis2005} While suppression of the valence transition in pure cerium lies beyond the reach of laboratory accessible magnetic fields, those in YbInCu$_4$ and Ce$_{0.8}$La$_{0.1}$Th$_{0.1}$ (i.e. chemically substituted Ce) occur at temperatures that are sufficiently low for their valence transitions to have been driven to zero. Moreover, both have been shown to exhibit elliptical phase boundaries of the form given by Equation (\ref{elliptical})\cite{dzero2000,drymiotis2005} (which we reproduce in Figs.~\ref{elliptical}a and b), and these have been shown to be consistent with the theoretical predictions of a valence transition.\cite{dzero2000}

\section{results}
\subsection{Phase boundary}

The evidence supporting our key finding that the suppression of $T_{\rm o}$ under a magnetic field in URu$_2$Si$_2$ (Fig.~\ref{elliptical}c) has the elliptical form found in valence transition systems is presented in Fig.~\ref{elliptical}. Crucial to our understanding the form of the phase boundary in URu$_2$Si$_2$ are the prior findings that the Sommerfeld coefficient and spin susceptibility are both strongly enhanced at temperatures both above and below the transition.\cite{palstra1985,maple1986,lopez1990,harrison2003,altarawneh2011,altarawneh2012} 
Thus, whereas YbInCu$_4$ and Ce$_{0.8}$La$_{0.1}$Th$_{0.1}$ involve a transition from local moment-like behavior at high temperatures to Fermi liquid-like behavior at low temperatures, URu$_2$Si$_2$ can be regarded as a Fermi liquid throughout. Under such circumstances, we can write approximate free energies of the form
\begin{eqnarray}\label{freeenergy}
F_{\rm U}=-\frac{1}{2}\gamma_{\rm U}T^2-\frac{1}{2\mu_0}(\chi_{\rm U}+\chi_{\rm bg})B^2+F_{\rm ph}~~~~~~~~\\\nonumber
F_{\rm L}=-\frac{1}{2}\gamma_{\rm L}T^2-\frac{1}{2\mu_0}(\chi_{\rm L}+\chi_{\rm bg})B^2+F_{\rm ph} -E_{\rm o},
\end{eqnarray}
for the upper (U) and lower (L) phases, where $\gamma_{\rm U,L}=\frac{\pi^2}{3}k^2_{\rm B}D_{\rm U,L}$ and $\chi_{\rm U,L}=\mu_0[g^\ast_{\rm eff}\sigma]^2\mu_{\rm B}^2D_{\rm U,L}$ refer to the Sommerfeld coefficients and spin susceptibilities in each of the phases,\cite{ashcroft1976} 
and where $g^\ast_{\rm eff}$ represents an effective {\it g}-factor for pseudospins of spin $\pm\sigma$.\cite{altarawneh2011,altarawneh2012}
Meanwhile, $\chi_{\rm bg}$ represents an additional background contribution to the magnetic susceptibility not arising from itinerant states (such as that arising from Van Vleck paramagnetism),  while $E_{\rm o}$ represents the enthalpy change associated with a change in state at the valence transition, analogous to that of a non symmetry-breaking liquid-gas transition. On defining the phase boundary as the line at which $F_{\rm U}=F_{\rm L}$, and neglecting possible changes in the phonon contribution $F_{\rm ph}$ across the valence transition, we arrive at the form of the phase boundary given by Equation (\ref{elliptical}), where $E_{\rm o}~=~\frac{\Delta\gamma}{2}T^2_{\rm o}~\equiv~\frac{\Delta\chi}{2\mu_0}B_{\rm o}^2$ and $\Delta\gamma=\gamma_{\rm U}-\gamma_{\rm L}$ and $\Delta\chi=\chi_{\rm U}-\chi_{\rm L}$ refer to changes in the Sommerfeld coefficient and spin susceptibility across the transition. 

One important characteristic of a purely entropy-driven elliptical phase boundary is that, regardless of specific values of $\Delta\gamma$ and $\Delta\chi$, the ratio, $T_{\rm o}/B_{\rm o}$, of the transition temperature to the strength of the critical magnetic field depends solely on the effective quasipaticle {\it g}-factor
\begin{equation}\label{gfactor}
g^\ast_{\rm eff}=\frac{\pi k_{\rm B}}{\sqrt{3}\sigma\mu_{\rm B}}~\frac{T_{\rm o}}{B_{\rm o}}.
\end{equation}
Since URu$_2$Si$_2$ can be regarded as a Fermi liquid on both sides of the phase transition,\cite{palstra1985,maple1986} Equation (\ref{gfactor}) is made more precise by the fact that $\Delta\gamma$ and $\Delta\chi$ are both defined in terms of the electronic density of states $D_{\rm U,L}$ in each of the phases.\cite{ashcroft1976} Because the effective {\it g}-factor of URu$_2$Si$_2$ has already been measured independently by different experimental Appendix,\cite{altarawneh2011,altarawneh2012} it provides us with a more quantitatively robust verification of a valence transition than what has been possible in other systems. On inserting the experimental values of $T_{\rm o}$ and $B_{\rm o}=$~35~T into Equation (\ref{gfactor}) for $\sigma=\frac{1}{2}$ pseudospins, we obtain $g^\ast_{\rm eff}\approx$~2.70, which is in excellent quantitative agreement with $g^\ast_{\rm eff}\approx$~2.6 from the measurements of spin zeroes in de~Haas-van~Alphen effect\cite{altarawneh2011} and $g^\ast_{\rm eff}\approx$~2.65 from the Pauli limited superconducting upper critical field.\cite{altarawneh2012} 

The {\it g}-factor analysis can also be extended to tilted magnetic fields, where the Ising anisotropy of the Zeeman splitting has been shown to 
cause the effective {\it g}-factor to acquire an angular dependence of the form $g^\ast_{\rm eff,\theta}=g^\ast_{\rm eff}\cos\theta$, where $\theta$ is angle by which the magnetic field is tilted away from the crystalline $\hat{c}$ axis. Inclusion of this Ising anisotropy into Equation (\ref{freeenergy}) causes the critical magnetic field to scale as $B_{\rm o,\theta}=B_{\rm o}/\cos\theta$, which agrees with the results of tilted high magnetic field experiments,\cite{jo2007} thus providing us with a further validation of the valence transition model.

The elliptical form of the phase boundary is largely conditional upon a broken symmetry, if present, having a negligible impact on the thermodynamics of the phase transition. Indeed, Fe doping-dependent studies have shown that the phase boundary departs from the elliptical form for the antiferromagnetic phase.\cite{ran2017} We take this argument further in Fig.~\ref{elliptical}d by contrasting the elliptical form of the phase boundary with those predicted for various forms of broken-symmetry phase. The difference in the form of the valence transition phase boundary from that of broken-symmetry phases becomes greater as the transition temperature is suppressed to zero by a magnetic field. On plotting its phase boundary in $T^2$-versus-$B^2$ coordinates in Fig.~\ref{elliptical}d, the squared transition temperature of a charge density-wave,\cite{mckenzie1997} or any other order parameter involving pairing between spin-up and spin-down states on the Fermi surface,\cite{maki1964} is generally expected to exhibit a downward curvature with increasing $B^2$, eventually giving rise to a precipitous drop of the phase boundary at a critical magnetic field\cite{maki1964} or a transition into an inhomogeneous phase with a substantially reduced transition temperature.\cite{mckenzie1997} In the case of a simple antiferromagnetic state,\cite{moll2017} by contrast, the squared transition temperature is typically found to exhibit an upward curvature with increasing $B^2$.

\subsection{Phase transition}

\begin{figure}[!!!!!!!htbp]
\centering 
\includegraphics*[width=.4\textwidth]{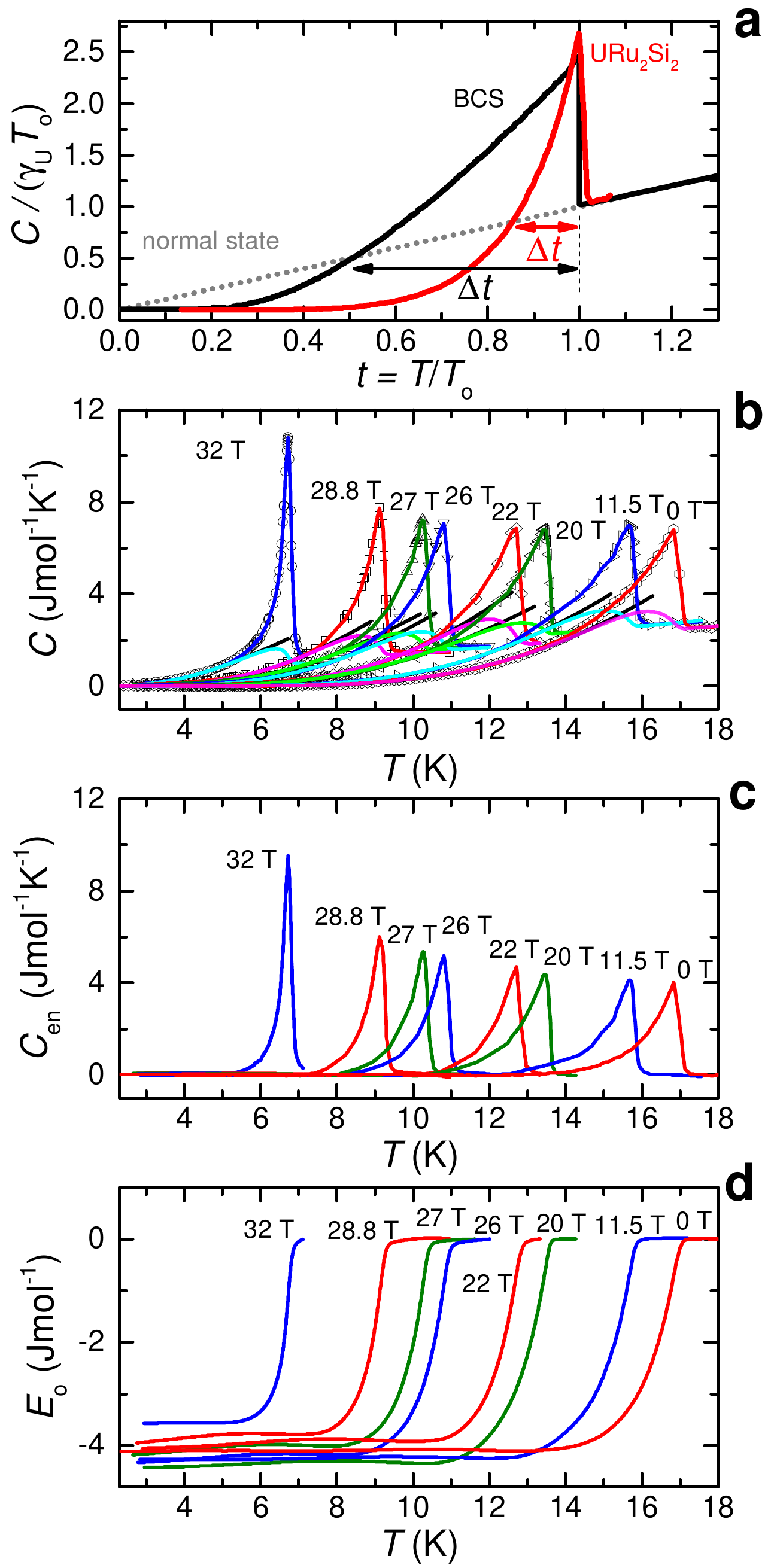}
\caption{{\bf a}, Comparison of the heat capacity transition observed in URu$_2$Si$_2$ ($B=$~0~T), with a BCS phase transition (black)\cite{johnston2013} and the normal state heat capacity (grey), plotted versus reduced temperature $t=T/T_{\rm o}$. In URu$_2$Si$_2$, {\it only} the phonon contribution and the low temperature electronic term $\gamma_{\rm L}T$ have been subtracted (see Appendix).\cite{vandijk1994,moriya1994} A salient feature of the anomaly, is the exponential behavior $C(t)\propto e^{-\frac{\Delta}{T}}$ for $t<1$. {\bf b}, The electronic heat capacity of URu$_2$Si$_2$ (data points from Ref.\cite{jaime2002,kim2003}) at different externally applied values of the magnetic field, as indicated. Alternating blue, red and olive lines represent interpolations of $C(T)$ data points (black symbols), while alternating cyan, magenta and green lines depict the forms of $C_{\rm cr}(T)$ extracted iteratively using Equation (\ref{heatcapacityform}), as described in the text. Exponential fits of $Ae^{-\frac{\Delta}{T}}$ to the measured $C(T)$ are made up to a temperature $T_{\rm max}<T_{\rm o}$ (black lines) for each magnetic field, where $T_{\rm max}$ is the temperature above which the measured $C(T)$ is observed to depart from a simple exponential form (see Appendix). To show the degree of departure, the fitted lines are extrapolated beyond $T_{\rm max}$. {\bf c}, Extracted forms of $C_{\rm en}(T)=C(T)-C_{\rm cr}(T)$ depicted in alternating blue, red and olive lines. {\bf d}, $\int_0^\infty C_{\rm en}{\rm d}T$ showing the experimentally estimated enthalpy change $E_{\rm o}$ at the transition.}
\label{heatcapacity}
\end{figure}

Further evidence supporting a valence transition is found in the form of the phase transition in heat capacity (plotted versus reduced temperature $t=T/T_{\rm o}$ in Fig.~\ref{heatcapacity}a),\cite{jaime2002} which we find to be appreciably sharper than that of a BCS phase transition.\cite{tinkham1996,gruner1994,johnston2013} Once the quasiparticle excitations giving rise to the exponential tail ($C(T)\propto e^{-\frac{\Delta}{T}}$ for $T<T_{\rm o}$)\cite{maple1986} are attributed to a hybridization gap opening between conduction and $f$ electron states\cite{schmidt2010,aynajian2010,park2012} rather than a Fermi surface gapped by an order parameter, the heat capacity loses any resemblance it might have had to a BCS phase transition. 

In general, when a BCS phase transition occurs, most of the enthalpy change associated with the growth of the order parameter occurs between $t=1$ and $t\approx$~0.5 (the point at which the heat capacity of the ordered phase has a value comparable to that $\gamma_{\rm U}T$ extrapolated from the normal state above the transition). Hence, the entropy loss resulting from the onset of a BCS order parameter is distributed over a broad range in reduced temperatures spanning $\Delta t\approx$~0.5 (see Fig.~\ref{heatcapacity}a),\cite{tinkham1996,gruner1994} and this continues to be so even in the case of strong coupling.\cite{johnston2013} In URu$_2$Si$_2$ (see Fig.~\ref{heatcapacity}a), however, we find that the bulk of the enthalpy change is compressed within a significantly narrower $\Delta t\lesssim$~0.14 range in reduced temperature. 

A crucial clue as to the origin of the sharp transition in URu$_2$Si$_2$ is revealed upon extending the range in temperature compared to earlier studies\cite{vandijk1994,maple1986} over which $C=Ae^{-\frac{\Delta}{T}}$ is fit to the $T<T_{\rm o}$ tail of the heat capacity in Fig.~\ref{heatcapacity}b  (see Appendix and Fig.~\ref{arrhenius} for an Arrhenius plot). Upon extending the range of fitting to lower temperatures, an upward departure from the exponential form becomes apparent on the approach to $T_{\rm o}$, with the degree of departure becoming increasingly pronounced in stronger magnetic fields (see Fig.~\ref{heatcapacity}b). Given the sharpness of the phase transition in Fig.~\ref{heatcapacity}a and b, and the previously identified hybridization gap origin of the exponential tail below $T_{\rm o}$,\cite{schmidt2010,aynajian2010,park2012} we proceed to show that the change in electronic contribution to the heat capacity in URu$_2$Si$_2$ is consistent with the sum 
\begin{equation}\label{heatcapacityform}
C(T)=C_{\rm en}(T)+C_{\rm cr}(T)
\end{equation}
of two contributions. The first contribution $C_{\rm en}(T)=-\alpha~\frac{\partial{\rm f}(T)}{\partial T}$ represents the enthalpy change at the valence transition. The second contribution $C_{\rm cr}(T)\approx{\rm f}(T)Ae^{-\frac{\Delta}{T}}+[1-{\rm f}(T)]\Delta\gamma T$ accounts for the crossover between L and U phases or, alternatively, their coexistence over the narrow range of temperatures at the transition. The volume fractions of the L and U phases are therefore ${\rm f}(T)$ and $1-{\rm f}(T)$, respectively. Together, the two terms in Equation (\ref{heatcapacityform}) produce a differential equation, which we solve iteratively for ${\rm f}(T)$ at different constant values of the magnetic field. We then plot the corresponding temperature-dependent $C_{\rm cr}(T)$ and $C_{\rm en}(T)$ at each magnetic field in Figs.~\ref{heatcapacity}b and c, respectively. Finally, the enthalpy change on crossing the phase boundary is given by the integration $E_{\rm o}=\int_0^\infty C_{\rm en}(T){\rm d}T$ (see Fig.~\ref{heatcapacity}d). 

The form of $C_{\rm en}$ resembles first order phase transitions observed in solid state systems\cite{hardy2009} and is also similar in form to the sharp transition observed in the  thermal expansion coefficient,\cite{aoki2010,kambe2013,devisser1986} for which an exponential term due to quasiparticle excitations is largely absent. We find the enthalpy change be approximately indpendent of magnetic field, which is consistent with the constant $E_{\rm o}$ assumed in Equation (\ref{freeenergy}). Meanwhile, $C_{\rm cr}(T)$ is found to be similar in shape to the transitions observed in the electrical resistivity\cite{zhu2009} and thermal conductivity,\cite{sharma2006} which we understand to be the consequence of electrical transport coefficients being insensitive to the enthalpy change at a transition.

\section{discussion}

The key experimental evidence supporting a valence transition in  URu$_2$Si$_2$ is the elliptical form of the phase boundary (see Fig.~\ref{elliptical}). We establish thermodynamic consistency with an entropy-driven valence transition by finding that the ratio of the critical magnetic field to the critical temperature is defined solely in terms of a quasiparticle {\it g}-factor in Equation (\ref{gfactor}) that is in excellent agreement with that previously obtained from other experiments.\cite{altarawneh2011,altarawneh2012} We further show that, despite the shape of the transition in the heat capacity being reported to resemble a BCS phase transition,\cite{palstra1985,maple1986} it is found to be considerably sharper upon making a direct comparison, and departs significantly from the standard form of a second order phase transition after taking into consideration the hybridization gap-origin of the the exponential tail observed in the heat capacity.\cite{schmidt2010,aynajian2010,park2012} 

There are several ways that the sharp transition in the heat capacity in Fig.~\ref{heatcapacity}b can be interpreted. The degree to which $f$-electrons are hybridized with conduction electrons is generally found to increase as the temperature is lowered in a mixed valence system, giving rise to a gradual change in the valence and the core $f$ electron occupancy with decreasing temperature. A valence transition occurs when the change in the strength of the hybridization, or effective Kondo temperature, becomes a non-linear function of the temperature.\cite{lawrence1981,dzero2000}

One recently proposed scenario is that the hybridization {\it is} an order parameter, giving rise to a second order phase transition in which the hybridization increases continuously from zero at $T=T_{\rm o}$.\cite{dubi2011,chandra2013} Since the enthalpy change at the transition in Fig.~\ref{heatcapacity}c is related to the strength of the hybridization, it provides a measure of how quickly such an order parameter must onset in temperature. The sharp change in enthalpy with temperature in Fig.~\ref{heatcapacity}c requires the order parameter to reach saturation within 1 or 2~K of the phase boundary, which is an indication of either strong coupling or a tendency to become first order. Further evidence for the hybridization gap reaching saturation quickly is provided by the ability of a simple exponential function to fit the tail of $C(T)$ over a broad range of $T$ (see Figs.~\ref{heatcapacity}b and \ref{arrhenius}), up to and including temperatures within 1 to 2~K of the phase boundary. One difficulty with the presently proposed hybridization order parameter scenarios, however, is that they require the onset of a subsidiary broken translational symmetry at $T_{\rm o}$, such as a charge density-wave accompanying hybridization density-wave\cite{dubi2011} or a spin-density wave accompanying hastatic order,\cite{chandra2013} neither of which have been detected.\cite{mydosh2011} 

An alternative possibility is that the strength of the hybridization (or the effective Kondo temperature) in URu$_2$Si$_2$ increases discontinuously,\cite{lawrence1981,dzero2000} giving rise to a first order phase transition of the type observed in YbInCu$_4$ or Ce$_{0.8}$La$_{0.1}$Th$_{0.1}$.  One benefit of postulating such a scenario in URu$_2$Si$_2$ is that it dispenses with need to identify hidden order parameter, which, as already discussed, has remained a major point of contention in this material.\cite{mydosh2011} First order phase transitions are confirmed in YbInCu$_4$ and Ce$_{0.8}$La$_{0.1}$Th$_{0.1}$ by the observation of sharp phase transitions and significant irreversible (i.e. hysteretic) contributions to the heat capacity.\cite{immer1997,drymiotis2005,silhanek2006} While the width of the phase transition in the heat capacity of URu$_2$Si$_2$ (in Fig.~\ref{heatcapacity}b) is similar to that observed in YbInCu$_4$,\cite{silhanek2006} a significant irreversible component of the heat capacity has yet to be reported in URu$_2$Si$_2$. On the other hand, a $\delta B\approx$~0.3\% hysteresis in $B_{\rm o}$ between rising and falling magnetic fields is observed,\cite{kim2003i} which, while small, provides a robust indication of a first order phase transition at $B=B_{\rm o}$. Hysteresis is also reported to accompany a small lattice distortion at $T_{\rm o}$, which is suggestive of a weakly first order phase transition at $B=0$.\cite{tonegawa2014,tabata2014} 

A possible variation of the above scenario, which would likely alter the dynamics of a first order phase transition, is that the valence change results from the passage of a hybridized $f$ electron band though the chemical potential.\cite{park2012,santandersyro2009} While the location of the hybridized bands in energy must ultimately be determined by the strength of the hybridization, such a scenario could result in a valence changing phase transition in the absence of a sharp discontinuity in the strength of the hybridization. For such a scenario, $\Delta$ (see Appendix) would no longer provide a direct measure of the hybridization gap, but instead provide a measure of distance of the narrow $f$ electron-like feature from the chemical potential. Possible support for this scenario is provided by observation of a Schotte-Schotte anomaly in strong magnetic fields (see Appendix).\cite{silhanek2005}

Should a valence transition occur in URu$_2$Si$_2$, the tetragonal crystal structure\cite{palstra1985,maple1986} is another factor having the potential to cause differences in its behavior compared to the well studied valence transitions in cubic materials.\cite{immer1997,lawrence1981} For example, electronic anisotropy could be an important factor in causing the hybridization gap within the L phase to be smaller and therefore more prone to quasiparticle excitations than those typically found in valence transition systems,\cite{palstra1985,maple1986} or in causing the transition to primarily involve a large change in the crystallographic anisotropy rather than the volume.\cite{devisser1986} Superconductivity, which occurs at temperatures below 2~K within the L phase of URu$_2$Si$_2$, can also be sensitive to the electronic anisotropy.\cite{palstra1985,maple1986} The absence of a hidden order parameter, if verified, would liberate superconductivity from the need to coexist with an unconventional broken-symmetry phase. It would also have the effect of reducing its phase diagram under pressure (within the L phase) to a simple competition between superconductivity and antiferromagnetism (which occurs under pressure in URu$_2$Si$_2$).\cite{villaume2008}  

Given that several key experimental features are consistent with the occurrence of an entropy-driven valence transition in URu$_2$Si$_2$, we conclude that the continued search for a hidden order parameter involving {\it either} localized {\it or} itenerant $f$ electron states is rendered largely unnecessary. While our findings do not preclude nucleation of a form of broken-symmetry at $T_{\rm o}$ as a subsidiary effect, such order, if present, must involve a change in energy that is too small to cause a discernible departure from the ideal elliptical form of the phase boundary expected for an entropy-driven valence transition.\cite{dzero2000} There is also clearly a need to confirm prior reports of weak hysteresis occurring in physical properties at or below $T_{\rm o}$ (at $B=$~0).\cite{tonegawa2014,tabata2014,park1998,schemm2015} 

The experimental and modeling work was supported by the US Department of Energy ``Science of 100 tesla" BES program. 
The original motivation for this work was developed as part of a Los Alamos National Laboratory LDRD program. We thank Piers Coleman, Premi Chandra and Brian Maple for insightful discussions.

\section{Appendix}

\subsection{Background subtraction in heat capacity measurements}
\begin{figure}[!!!!!!!htbp]
\centering 
\includegraphics*[width=.45\textwidth]{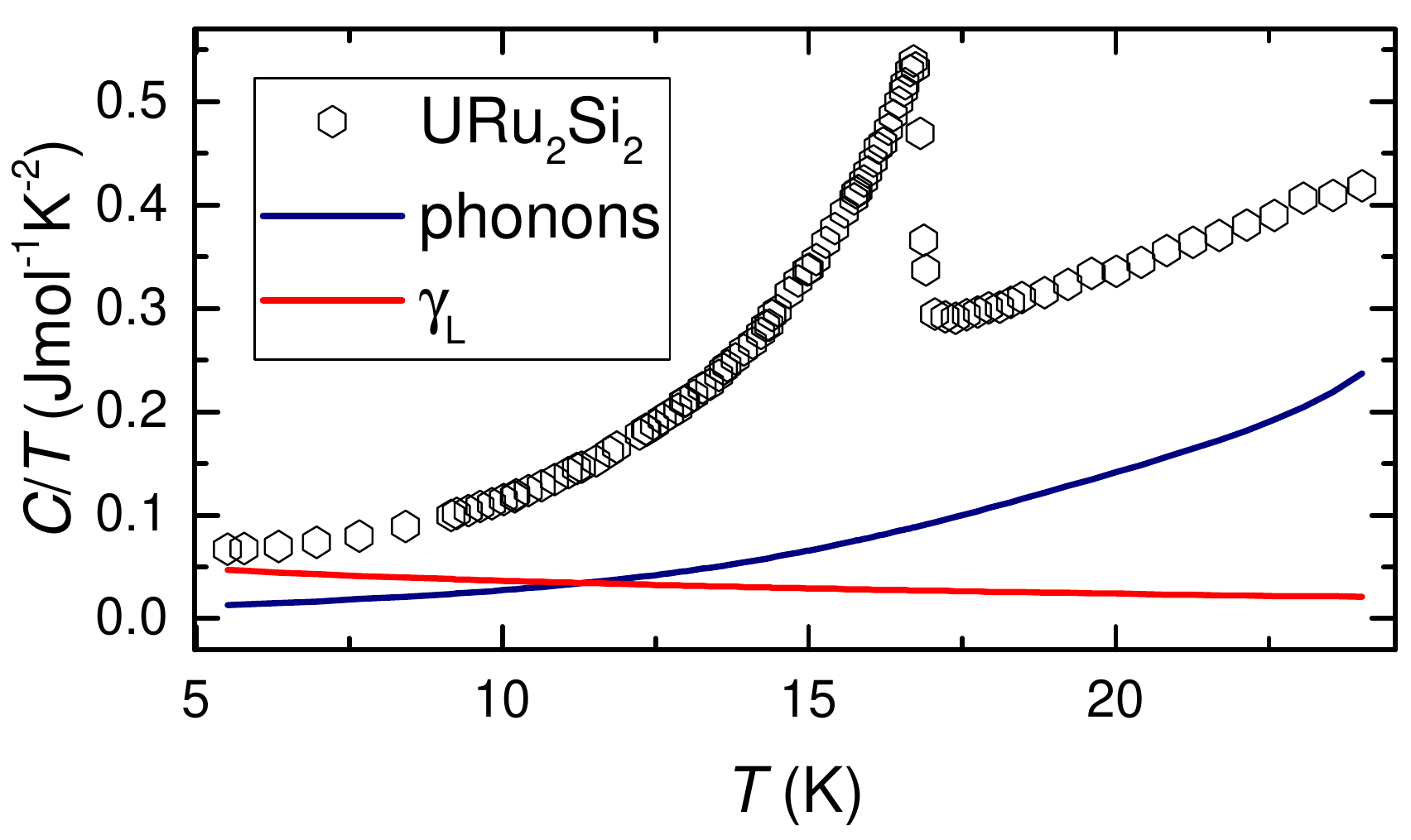}
\caption{Raw heat capacity of URu$_2$Si$_2$ at $B=0$\cite{jaime2002}, with the phonon (i.e. ThRu$_2$Si$_2$) and spin fluctuation\cite{moriya1994,vandijk1994} components indicated.}
\label{subtraction}
\end{figure}

Following the procedure outlined in previously published heat capacity measurements,\cite{vandijk1994,jaime2002} the phonon contribution is obtained from measurements of ThRu$_2$Si$_2$ (see Fig.~\ref{subtraction}). After subtraction, the remaining heat capacity is assumed to be electronic in origin. At $B=0$, the exponential contribution to the heat capacity is observed to extend down to $\approx$~6~K, below which $C(T)\approx\gamma_{\rm L}T$. 

In order to extract the form of the phase transition in the heat capacity, the low temperature Sommerfeld contribution needs to be subtracted. This was initially done by subtracting a constant $\gamma_{\rm L}$ from the measured heat capacity divided by temperature i.e. ($C/T$).\cite{maple1986} More recent heat capacity studies, have shown that $\gamma_{\rm L}$ increases slowly with decreasing temperature, and Moriya and Takimoto\cite{moriya1994} and van~Dijk {\it et al.}\cite{vandijk1994} have shown that this behavior can be attributed to spin fluctuations, originating from the close proximity of URu$_2$Si$_2$ to an antiferromagnetic ground state. Throughout our manuscript we have assumed $\gamma_{\rm L}(T)$ to have the form determined by van~Dijk {\it et al.} (see Fig.~\ref{subtraction}).

\subsection{Determination of the exponential tail in the heat capacity for $T<T_{\rm o}$}
\begin{figure}[!!!!!!!htbp]
\centering 
\includegraphics*[width=.45\textwidth]{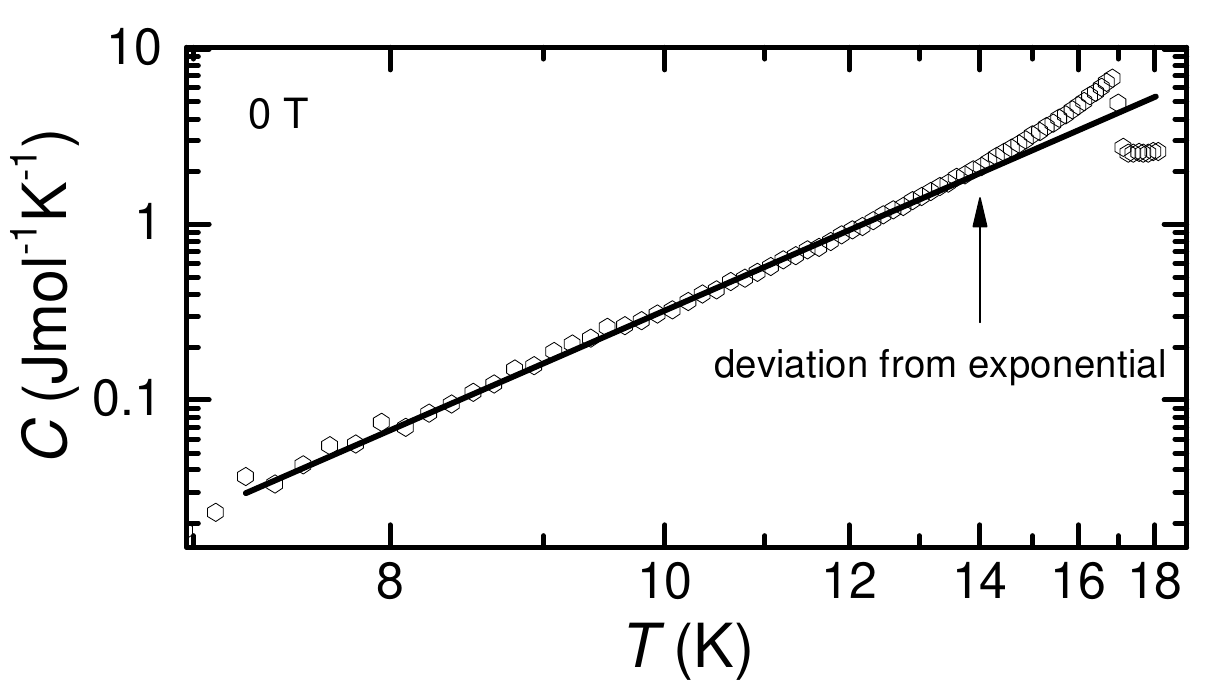}
\caption{An Arrhenius plot of the heat capacity of URu$_2$Si$_2$ at $B=0$, after subtraction of the phonon and spin fluctuation components, so as to show the departure from exponential behavior on the approach to $T_{\rm o}$ on increasing the temperature.}
\label{arrhenius}
\end{figure}

A striking feature of the experimental heat capacity in URu$_2$Si$_2$ in strong magnetic fields, is the departure from simply exponential behavior $C(T)\propto e^{-\frac{\Delta}{T}}$ as the temperature is increased towards $T_{\rm o}$. A closer examination of the heat capacity reveals such a departure also occurs at $B=0$, which becomes more clear on constructing an Arrhenius plot (see Fig.~\ref{arrhenius}). To determine the point of departure, we begin by fitting an exponential curve to the entire region below $T_{\rm o}$ and then repeat the fitting while incrementally reducing the upper limit of the fit, $T_{\rm max}$. We determine the optimal $T_{\rm max}$ as that below which the exponential fit is no longer observed to change on reducing the range. This procedure is repeated at all magnetic fields.

\subsection{Estimates of the hybridization gap $\Delta$}
\begin{figure}[!!!!!!!htbp]
\centering 
\includegraphics*[width=.35\textwidth]{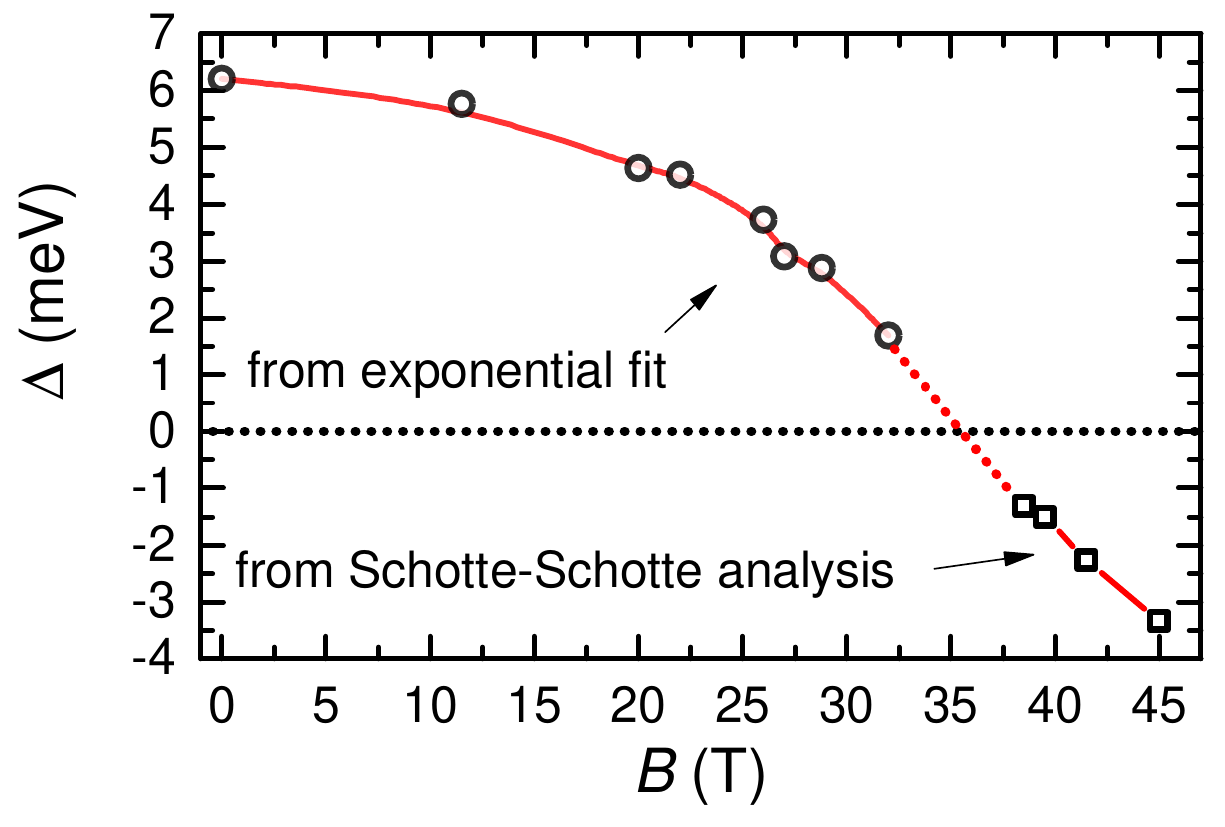}
\caption{Magnetic field-dependence of the gap value $\Delta$ (circles) obtained by fitting $Ae^{-\frac{\Delta}{T}}$ to $C(T)$ below the phase transition in Fig.~\ref{heatcapacity}b, as described in the text. Also shown, is the $5f$ electron level (squares) relative to the chemical potential, inferred from the Schotte-Schotte anomaly observed in strong magnetic fields.\cite{silhanek2005} The red dotted line connecting the datasets is a guide to the eye.}
\label{order}
\end{figure}

Figure~\ref{order} shows $\Delta(B)$ obtained from fitting $Ae^{-\frac{\Delta}{T}}$ to the measured $C(T)$ up to a temperature $T_{\rm max}<T_{\rm o}$, as described in Fig.~\ref{heatcapacity}b.The magnitude of the rate $\frac{\partial\Delta}{\partial B}\approx$~$-$~0.4~meV per T at which $\Delta(B)$ falls to zero with increasing magnetic field is similar to the rate at which the $f$ electron levels are observed to move away from the chemical potential in very strong magnetic field heat capacity measurements (also plotted in Fig.~\ref{order}).\cite{silhanek2005} A scenario in which the hybridized $f$ electron band passes through the chemical potential at $B=B_{\rm o}$ to emerge at the other side of the chemical potential at $B>B_{\rm o}$ is therefore suggested. Point contact spectroscopy measurements\cite{park2012} and the observations of metamagnetism\cite{harrison2003} and light quasiparticles\cite{harrison2013} in strong magnetic fields suggest that while the hybridization in URu$_2$Si$_2$ does eventually vanish, this occurs outside the L phase.

\subsection{Robustness of the free energy}
\begin{figure}[!!!!!!!htbp]
\centering 
\includegraphics*[width=.45\textwidth]{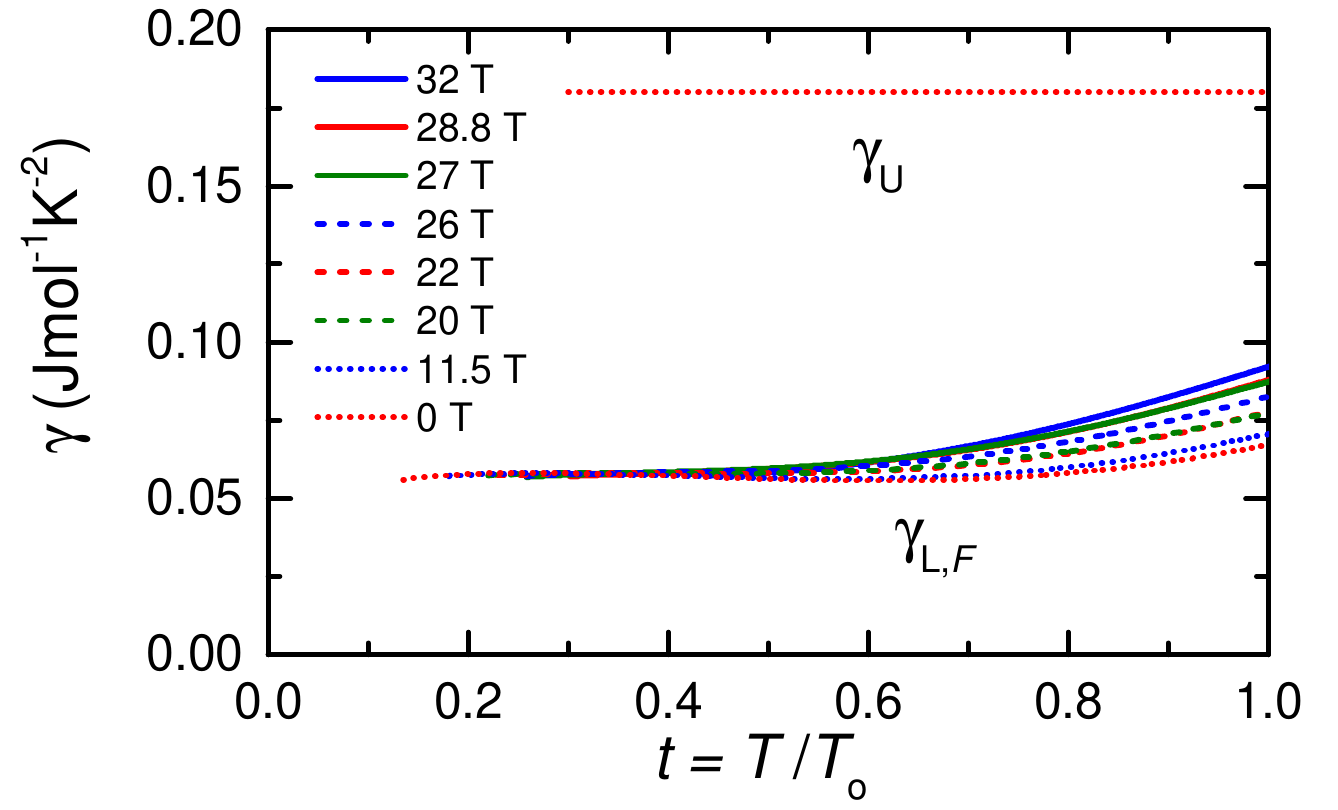}
\caption{The modified Sommerfeld coefficient $\gamma_{{\rm L},F}(T)$to be used in Equation (\ref{freeenergy}), as described in the text, obtained at different values of the magnetic field using Equation (\ref{enthalpy}). $\gamma_{\rm U}$ is also shown (at $B=0$), whose temperature and magnetic field-dependences remain undetermined.}
\label{effectivegamma}
\end{figure}

While Equation (\ref{freeenergy}) assumes a simple quadratic form for the temperature-dependence of the free energy and magnetic field-dependence of the susceptibility, departures are expected due to the spin fluctuations,\cite{moriya1994,vandijk1994} quasiparticle excitations cross the hybridization gap and non-linearities in the susceptibility with increasing magnetic field. Under such circumstances, 
we must instead use
\begin{eqnarray}\label{enthalpy}
\gamma_{{\rm L},F}(T,B)=\frac{2}{T^2}\int_0\int_0\bigg(\gamma_{\rm L}(T)+\frac{C_{\rm cr}(T,B)}{T}\bigg){\rm d}T^2\\
\chi_{{\rm L},F}(T,B)=\frac{2\mu_0}{B^2}\int_0M(T,B){\rm d}B~~~~~~~~~~~~~~~~~~~
\end{eqnarray}
in place of $\gamma_{\rm L}$ and $\chi_{\rm L}$ in Equation (\ref{freeenergy}). In Fig.~\ref{effectivegamma}, $\gamma_{{\rm L},F}(T)$ is found to be very weakly dependent on temperature at $B=0$, but appears to exhibit an upturn with magnetic field at finite temperatures. It nevertheless remains significantly smaller than $\gamma_{\rm U}$. Prior measurements of the magnetic susceptibility show that $M(B)$ also undergoes an upturn in advance of the phase boundary when $T=$~7 and 8 K, suggesting a small increase in $\chi_{{\rm L},F}(B)$.\cite{harrison2003} While $\chi\propto\gamma$, it has yet to be determined whether and to what extent $\chi_{{\rm L},F}(T)\propto \gamma_{{\rm L},F}(T)$ in URu$_2$Si$_2$.

\begin{thebibliography}{99}

\bibitem{mydosh2011} Mydosh, J. A., Oppeneer, P. M., Hidden order, superconductivity, and magnetism: The unsolved case of URu$_2$Si$_2$. {\it Rev. Mod. Phys.} {\bf 83}, 1301-1322 (2011).

\bibitem{kasuya1997} Kasuya, T., Hidden ordering and heavy mass in URu$_2$Si$_2$ and its alloys. {\it J. Phys. Soc. Japan} {\bf 66}, 3348-3351 (1997).

\bibitem{chandra2002} Chandra, P., Coleman, P., Mydosh, J. A., Tripathi, V., Hidden orbital order in the heavy fermion metal URu$_2$Si$_2$. {\it Nature} {\bf 417}, 831-834 (2002).

\bibitem{palstra1985} Palstra, T. T. M., Menovsky, A. A., Vandenberg, J., Dirkmaat, A. J., Kes, P. H., Nieuwenhuys, G. J., Mydosh, J. A., Superconducting and magnetic transitions in the heavy-fermion system URu$_2$Si$_2$, {\it Phys. Rev. Lett.} {\bf 55}, 2727-2730 (1985).

\bibitem{maple1986} Maple, M. B., Chen, J. W., Dalchaouch, Y., Kohara, T., Rossel, C., Torikachvili, M. S., McElfresh, M. W., Thompson. J. D., Partially-gapped Fermi-surface in the heavy-electron superconductor URu$_2$Si$_2$. {\it Phys. Rev. Lett.} {\bf 56}, 185-188 (1986).


\bibitem{varma2006} Varma, C. M., Zhu, L. J., Helicity order: Hidden order parameter in URu$_2$Si$_2$. {\it Phys. Rev. Lett.} {\bf 96}, 036405 (2006).

\bibitem{elgazzar2009} Elgazzar, S., Rusz, J., Amft, M., Oppeneer, P. M.,  Mydosh, J. A., Hidden order in URu$_2$Si$_2$ originates from Fermi surface gapping induced by dynamic symmetry-breaking. {\it Nature Mat.} {\bf 8}, 337-341 (2009).

\bibitem{okazaki2011} Okazaki, R., Shibauchi, T., Shi, H. J., Haga, Y., Matsuda, T. D., Yamamoto, E., Onuki, Y., Ikeda, H., Matsuda, Y., Rotational symmetry-breaking in the Hidden-Order Phase of URu$_2$Si$_2$. {\it Science} {\bf 331}, 439-442 (2011).

\bibitem{fujimoto2011} Fujimoto, S., Spin nematic state as a candidate of the hidden order phase of URu$_2$Si$_2$. {\it Phys. Rev. Lett.} {\bf 106}, 196497 (2011).

\bibitem{pepin2011} Pepin, C., Norman, M. R., Burdin, S., Ferraz, A., Modulated Spin Liquid: A New Paradigm for URu$_2$Si$_2$. {\it Phys. Rev. Lett.} {\bf 106}, 106601 (2011).


\bibitem{dubi2011} Dubi, Y., Balatsky, A. V., Hybridization Wave as the``Hidden Order'' in URu$_2$Si$_2$. {\it Phys. Rev. Lett.} {\bf 106}, 086401 (2011).

\bibitem{chandra2013} Chandra, P., Coleman, P., Flint, R., Hastatic order in the heavy-fermion compound URu$_2$Si$_2$. {\it Nature} {\bf 493}, 621-626 (2013).


\bibitem{santini1994} Santini, P., Amoretti, G., Crystal-field model of the magnetic-properties of URu$_2$Si$_2$, {\it Phys. Rev. Lett.} {\bf 73}, 1027-1030 (1994).

\bibitem{barzykin1995} Barzykin, V., Gor'kov, L. P., Singlet magnetism in heavy fermions. {\it Phys. Rev. Lett.} {\bf 74}, 4301-4304 (1995). 

\bibitem{ikeda1998} Ikeda, H., Ohashi, Y., Theory of unconventional spin density wave: A possible mechanism of the micromagnetism in U-based heavy fermion compounds. {\it Phys. Rev. Lett.} {\bf 81}, 3723-3726 (1998).

\bibitem{ohkawa1999} Ohkawa, F. J., Shimizu, H., Quadrupole and dipole orders in URu$_2$Si$_2$. {\it J. Phys.-Cond. Matt.} {\bf 11}, L519-L524 (1999). 

\bibitem{kiss2005} Kiss, A., Fazekas, P., Group theory and octupolar order in URu$_2$Si$_2$. {\it Phys. Rev. B} {\bf 71}, 054415 (2005). 

\bibitem{haule2009} Haule, K., Kotliar, G., Arrested Kondo effect and hidden order in URu$_2$Si$_2$. {\it Nature Phys.} {\bf 5}, 796-799 (2009).

\bibitem{harima2010} Harima, H., Miyake, K., Flouquet, J., Why the hidden order in URu$_2$Si$_2$ Is still hidden -- one simple answer. {\it J. Phys. Soc/ japan} {\bf 79}, 033705 (2010). 

\bibitem{kusunose2011} Kusunose, H., Harima, H., On the hidden order in URu$_2$Si$_2$ -- antiferro hexadecapole order and its consequences. {\it J. Phys. Soc. Japan} {\bf 80}, 084702 (2011).

\bibitem{ikeda2012} Ikeda, H., Suzuki, M. T., Arita, R., Takimoto, T., Shibauchi, T., Matsuda, Y., Emergent rank-5 nematic order in URu$_2$Si$_2$. {\it Nature Phys.} {\bf 8}, 528-533 (2012).

\bibitem{kung2015} Kung, H. H., Baumbach, R. E., Bauer, E. D., Thorsmolle, V. K., Zhang, W. L., Haule, K., Mydosh, J. A., Blumberg, G., Chirality density wave of the``hidden order'' phase in URu$_2$Si$_2$. {\it Science} {\bf 347}, 1339-1342 (2015). 



\bibitem{jeffries2010} Jeffries, J. R., Moore, K. T., Butch, N. P., Maple, M. B., Degree of 5$f$ electron localization in URu$_2$Si$_2$: Electron energy-loss spectroscopy and spin-orbit sum rule analysis. {\it Phys. Rev. B} {\bf 82}, 033103 (2010).

\bibitem{booth2016} Booth, C. H., Medling, S. A., Tobin, J. G., Baumbach, R. E., Bauer, E. D., Sokaras, D., Nordlund, D., Weng, T.-C., Probing 5$f$-state configurations in URu$_2$Si$_2$ with U $L_{III}$-edge resonant x-ray emission spectroscopy. {\it Phys. Rev. B} {\bf 94}, 045121 (2016).



\bibitem{dzero2000} Dzero, M. O., Gor'kov, L. P., Zvezdin, A. K., First-order valence transition in YbInCu$_4$ in the ($B$, $T$)-plane. {\it J. Phys.: Condens. Matter} {\bf 12}, L711-L718 (2000). 

\bibitem{immer1997} Immer, C. D., Sarrao, J. L., Fisk, Z. Magnetic-field, pressure, and temperature scaling of the first-order valence transition in pure and doped YbInCu$_4$. {\it Phys. Rev. B} {\bf 56}, 71-74 (1997).

\bibitem{drymiotis2005} Drymiotis, F., Singleton, J., Harrison, N., Lashley, J. C., Bangura, A., Mielke, C. H., Balicas, L., Fisk, Z., Migliori, A., Smith, J. L., Suppression of the $\gamma$-$\alpha$ structural phase transition in Ce$_{0.8}$La$_{0.1}$Th$_{0.1}$ by large magnetic fields. {\it J. Phys.: Condes, Matter} {\bf 17}, L77-L83 (2005).


\bibitem{jaime2002} Jaime, M., Kim, K. H., Jorge, G., McCall, S., Mydosh, J. A., High magnetic field studies of the hidden order transition in URu$_2$Si$_2$. {\it Phys. Rev. Lett.} {\bf 89}, 287201 (2002). 

\bibitem{kim2003} Kim, J. S., Hall, D., Kumar, P., Stewart, G. R., Specific heat of URu$_2$Si$_2$ in fields up to 42 T: Clues to the hidden order. {\it Phys. Rev. B} {\bf 67}, 014404 (2003).


\bibitem{schmidt2010} Schmidt, A. R., Hamidian, M. H., Wahl, P., Meier, F., Balatsky, A. V., Garrett, J. D., Williams, T. J., Luke, G. M., Davis, J. C., Imaging the Fano lattice to `hidden order' transition in URu$_2$Si$_2$. {\it Nature} {\bf 465}, 570-576 (2010).

\bibitem{aynajian2010} Aynajian, P., da Silva Neto, E. H., Parker, C. V., Huang, Y., Pasupathy, A., Mydosh, J. A., Yazdani, A., Visualizing the formation of the Kondo lattice and the hidden order in URu$_2$Si$_2$. {\it Proc. Nat. Acad. Sci. USA} {\bf 107}, 10383-10388 (2010). 

\bibitem{park2012} Park, W. K., Tobash, P. H., Ronning, F., Bauer, E. D., Sarrao, J. D., Thompson, J. D., Greene, L. H., Observation of the Hybridization Gap and Fano Resonance in the Kondo Lattice URu$_2$Si$_2$. {\it Phys. Rev. Lett.} {\bf 108}, 246403 (2012).

\bibitem{butch2015} Butch, N. P., Manley, M. E., Jeffries, J. R., Janoschek, M., Huang, K., Maple, M. B., Said, A. H., Leu, B. M., Lynn, J. W., Symmetry and correlations underlying hidden order in URu$_2$Si$_2$, {\it Phys. Rev. B} {\bf 91}, 035128 (2015).


\bibitem{park1998} Park, J.-G., Rib, H.-C., Kuwahara, K, Amitsuka, H., Thermal hysteresis in magnetization of single crystal URu$_2$Si$_2$.  {\it Journal of Magn. and Magn. Mater.} {\bf 177-181}, 455-456 (1998).

\bibitem{tonegawa2014} Tonegawa, S., Kasahara, S., Fukuda, T, Sugimoto, K., Yasuda, N., Tsuruhara1, Y., Watanabe, D., Mizukami, Y., Haga, Y., Matsuda. T. D., Yamamoto, E., Onuki, Y., Ikeda, H., Matsuda, Y., Shibauchi, T., Direct observation of lattice symmetry-breaking at the hidden-order transition in URu$_2$Si$_2$. {\it Nature Commun.} DOI: 10.1038/ncomms5188 (2014).

\bibitem{tabata2014} Tabata, C., Inami, T., Michimura, S., Yokoyama, M., Hidaka, H., Yanagisawa, T., Amitsuka, H. X-ray backscattering study of crystal lattice distortion in hidden order of URu$_2$Si$_2$. {\it Phil. Mag.}  {\bf 94}, 3691-3701 (2014).

\bibitem{schemm2015} Schemm, E. R., Baumbach, R. E., Tobash, P. H., Ronning, F., Bauer, E. D., Kapitulnik, A., Evidence for broken time-reversal symmetry in the superconducting phase of URu$_2$Si$_2$. {\it Phys. Rev. B} {\bf 91}, 140506 (2015).




\bibitem{lawrence1981} Lawrence, J. W., Riseborough, P., S., Parks, R. D., Valence fluctuation phenomena. {\it Rep. Prog. Phys.} {\bf 44}, 1-84 (1981). 


\bibitem{lopez1990} L\'{o}pez~de~la~Torre, M. A., Vieira, S., Villar, R., Maple, B. B., Torikachvili, M. S., Low temperature specific heat of URu$_2$Si$_2$ near the superconducting transition. {\it Physica B} {\bf 165 \& 166}, 385-386 (1990).

\bibitem{harrison2003} Harrison, N., Jaime, M., Mydosh, J. A., Reentrant hidden order at a metamagnetic quantum critical end point. {\it Phys. Rev. Lett.} {\bf 90}, 096402 (2003).

\bibitem{altarawneh2011} Altarawneh, M. M., Harrison, N., Sebastian, S. E., Balicas, L., Tobash, P. H., Thompson, J. D., Ronning, F.,  Bauer, E. D., Sequential spin polarization of the Fermi surface pockets in URu$_2$Si$_2$ and Its Implications for the hidden order. {\it Phys. Rev. Lett.} {\bf 106}, 146403 (2011).

\bibitem{altarawneh2012} Altarawneh, M. M., Harrison, N., Li, G., Balicas, L., Tobash, P. H., Ronning, F., Bauer, E. D., Superconducting pairs with extreme uniaxial anisotropy in URu$_2$Si$_2$. {\it Phys. Rev. Lett.} {\bf 108}, 066407 (2012).

\bibitem{ashcroft1976} Ashcroft, N. W., Mermin, N. D. Solid state physics (Saunders College Publishing, Orlando 1976).

\bibitem{jo2007} Jo, Y. J., Balicas, L.,Capan, C., Behnia, K.,Lejay, P., Flouquet, J., Mydosh, J. A., Schlottmann, P., Field-Induced Fermi surface reconstruction and adiabatic continuity between antiferromagnetism and the hidden-order state in URu$_2$Si$_2$. {\it Phys. Rev. Lett.} {\bf 98}, 166404 (2007).

\bibitem{ran2017} Ran, S., Jeon, I., Pouse, N., Breindel, A. J., Kanchanavatee, N., Huang, K., Gallagherd, A., Chen, K.-W., Graf, D., Baumbach, R. E., Singleton, J., Maple, M. B., Phase diagram of URu$_{2–x}$Fe$_x$Si$_2$ in high magnetic fields. {\it Proc. Nat. Acad. Sci. USA} {\bf 114}, 9826-9831 (2017).



\bibitem{mckenzie1997} McKenzie, R. H., Is the ground state of $\alpha$-(BEDT-TTF)$_2$MHg(SCN)$_4$ [$M=$~K,Rb,Tl]
a charge-density wave or a spin-density wave? preprint: arXiv:cond-mat/9706235v2.

\bibitem{maki1964}  Maki. K., Toshihito, T., Pauli Paramagnetism and Superconducting State. {\it Prog. Theor. Phys.} {\bf 31}, 954-956 (1964).

\bibitem{moll2017} Moll, P. J. W., Helm, T., Zhang, S.-S., Batista, C. D., Harrison, N., McDonald, R. D., Winter, L. E., Ramshaw, B. J., Chan, M. K., Balakirev, F. F., Batlogg, B., Bauer, E. D., Ronning, F.,Emergent magnetic anisotropy in the cubic heavy-fermion metal CeIn$_3$, {Quantum Materials} {\bf 2}, 46 (2017).


\bibitem{tinkham1996} Tinkham, M., Introduction to superconductivity (McGraw-Hill, Inc., New York, 1996).

\bibitem{gruner1994} G.~Gr\"{u}ner, Density Waves in Solids (Addison-Wesley, 1994).

\bibitem{johnston2013} Johnston, D. C., Elaboration of the $\alpha$-model derived from the BCS theory of superconductivity. {\it Supercond. Sci. Technol.} {\bf 26}, 115011 (2013).

\bibitem{vandijk1994} van~Dijk, N. H., Bourdarot, F., Klaasse, J. C. P., Hagmusa, I. H., Br\"{u}ck, E., Menovsky, A. A., Specific heat of heavy-fermion URu$_2$Si$_2$ in high magnetic fields, {\it Phys. Rev. B} {\bf 56}, 14493-14498 (1997).




\bibitem{hardy2009} Hardy, V., Br\'{e}ard, Y., Martin, C., Derivation of the heat capacity anomaly at a first-order transition by using a
semi-adiabatic relaxation technique, {\it J. Phys.: Condens. Matter} {\bf 21}, 075403 (2009).

\bibitem{devisser1986} de~Visser, A., Kayzel, F. E., Menovsky, A. A., Franse, J. J. M., Thermal expansion and specific heat of monocrystalline URu$_2$Si$_2$. {\it Phys. Rev. B} {\bf 34}, 8168-8171 (1986).

\bibitem{aoki2010} Aoki, D., Bourdarot, F., Hassinger, E., Knebel, G., Miyake, A., Raymond, S., Taufour, V., Flouquet, J., Field re-entrant hidden-order phase under pressure in URu$_2$Si$_2$, {\it J. Phys.: Condens. Matter} {\bf 22}, 164205 (2010).

\bibitem{kambe2013} Kambe, S., Aoki, D., Salce, B., Bourdarot, F., Braithwaite, D., Flouquet, J., Brison, J.-P., Thermal expansion under uniaxial pressure in URu$_2$Si$_2$, {\it Phys. Rev. B} {\bf 87}, 115123 (2013).


\bibitem{zhu2009} Zhu, Z., Hassinger, E., Xu, Z., Aoki, D., Flouquet, J., Behnia, K., Anisotropic inelastic scattering and its interplay with superconductivity in URu$_2$Si$_2$, {\it Phys. Rev. B} {\bf 80}, 175501 (2009).

\bibitem{sharma2006} Sharma, P. A., Harrison, N., Jaime, M., Oh,  Y. S., Kim, K. H., Batista, C. D., Amitsuka, H., Mydosh, J. A., Phonon Thermal transport of URu$_2$Si$_2$: Broken translational symmetry and strong-coupling of the ``hidden order'' to the lattice. {\it Phys. Rev. Lett.} {\bf 97}, 156401 (2006). 



\bibitem{silhanek2006} Silhanek, A. V., Jaime, M., Harrison, N., Fanelli, V. R., Batista, C. D., Amitsuka, H., Nakatsuji, S., Balicas, L., Kim, K. H., Fisk, Z., Sarrao, J. L., Civale, L., Mydosh, J. A., Irreversible dynamics of the phase boundary in U(Ru$_{0.96}$Rh$_{0.04}$)$_2$Si$_2$
and implications for ordering. {\it Phys. Rev. Lett.} {\bf 96}, 136403 (2006).






\bibitem{kim2003i} Kim, K. H., Harrison, N., Jaime, M., Boebinger, G. S., Mydosh, J. A., Magnetic-field-induced quantum critical point and competing order parameters in URu$_2$Si$_2$. {\it Phys. Rev. Lett.} {\bf 91}, 256401 (2003).









\bibitem{santandersyro2009} Santander-Syro, A. F., Klein, M., Boariu, F. L., Nuber, A., Lejay, P., Reinert, F., Fermi-surface instability at the `hidden-order' transition of URu$_2$Si$_2$. {\it Nature Phys.} {\bf 5}, 637-641 (2009).

\bibitem{silhanek2005} Silhanek, A. V., Harrison, N., Batista, C. D., Jaime, M., Lacerda, A., Amitsuka, H., Mydosh, J. A., Quantum critical $5f$ electrons avoid singularities in URu$_2$Si$_2$. {\it Phys. Rev. Lett.} {\bf 95}, 026403 (2005).



\bibitem{villaume2008} Villaume, A., Bourdarot, F., Hassinger, E., Raymond, S., Taufour, V., Aoki, D., Flouquet, J., Signature of hidden order in heavy fermion superconductor URu$_2$Si$_2$: Resonance at the wave vector ${\bf Q}=[1,0,0]$, {\it Phys. Rev. B} {\bf 78} 012204 (2008).


\bibitem{moriya1994} Moriya, T., Takimoto, T. Anomalous properties around magnetic instability in heavy electron systems. {\it J. Phys. Soc. Japan} {\bf 64}, 960-969 (1994).

\bibitem{harrison2013} Harrison, N., Moll, P. J. W., Sebastian, S. E., Balicas, L., Altarawneh, M. M., Zhu, J.-X., Tobash, P. H., Ronning, F., Bauer, E. D., Batlogg, B. Magnetic field-tuned localization of the $5f$-electrons in URu$_2$Si$_2$. {\it Phys. Rev. B} {\bf 88}, 241108 (2013).












\end{thebibliography}
\end{document}